\documentclass[
    ,final            
  ]
  {aipproc}
\usepackage{graphicx}
\usepackage{amsfonts}
\usepackage{amssymb}
\usepackage{amsmath}
\usepackage{xcolor}

\layoutstyle{6x9}

\newcommand{\lc}{{\cal L}}

\newcommand{\oh}{{1\over2}}
\newcommand{\trh}{{3\over2}}

\newcommand{\be}{\begin{eqnarray}}
\newcommand{\ee}{\end{eqnarray}}

\newcommand{\Eq}[1]{Eq.~(\ref{#1})}

\newcommand{\del}{\Delta}

\newcommand{\db}{\bar{D}}

\newcommand{\dvb}{\bar{D}^*}

\begin{document}

\title{Exotic dynamically generated baryons with C$=-$1}


\author{D.~Gamermann}{address={Departament d'Estructura i Constituents de la 
Mat\`eria and Institut de Ci\`encies del Cosmos,
Universitat de Barcelona, Avda. Diagonal 647, E-08028 Barcelona, Spain},
altaddress={Instituto de F\'isica corpuscular (IFIC), Centro Mixto
Universidad de Valencia-CSIC,\\ Institutos de Investigaci\'on de
Paterna, Aptdo. 22085, 46071, Valencia, Spain}}

\author{C.~Garc\'ia-Recio}{address={Departamento de F\'isica At\'omica, Molecular y 
Nuclear, Universidad de Granada, E-18071 Granada, Spain}}

\author{J.~Nieves}{address={Instituto de F\'isica corpuscular (IFIC), Centro Mixto
Universidad de Valencia-CSIC,\\ Institutos de Investigaci\'on de
Paterna, Aptdo. 22085, 46071, Valencia, Spain}}

\author{L.L.~Salcedo}{address={Departamento de F\'isica At\'omica, 
Molecular y Nuclear, Universidad de Granada, E-18071 Granada, Spain}}

\author{L.~Tolos}{address={Theory Group, KVI, University of Groningen, 
Zernikelaan 25, 9747 AA Groningen, The Netherlands},altaddress={Instituto de Ciencias del Espacio (IEEC/CSIC), Campus Universitat Autonoma de Barcelona,
Facultat de Ciencies, Torre C5, E-08193, Bellaterra (Barcelona), Spain}}

\begin{abstract}
 We follow a model based on the SU(8) symmetry for the interaction of mesons with baryons. The model treats on an equal footing the pseudo-scalars and the vector mesons, as required by heavy quark symmetry. The T-matrix calculated within an unitary scheme in coupled channels has poles which are interpreted as baryonic resonances.  
\end{abstract}

\maketitle


\section{Introduction}

There is a long standing discussion on the structure of certain hadrons. Although the ground state hadrons fit
quark model predictions, many excited states have properties which are no well described by these models. For these
hadrons, whose structure is not well described by quark models, one appealing possibility is that they are are dynamically
generated states. That means they are quasi bound states generated by the interaction of the ground state hadrons. 
This picture of dynamically generated states has been successful in describing numerous observed meson and baryon excited
states in the charmed and non-charmed sectors. 

In the present work we want to explore the possibility of generating dynamically exotic baryon states with negative charm
quantum number. The possible existence of such baryons has already been investigated in \cite{hofmann,hofmann2}, where
the interaction of pseudo scalar mesons with ground state $\oh^+$ and $\trh^+$ baryons have been studied and dynamically
generated states have been obtained for $C$=2, 1, 0 and -1. The meson baryon interaction in these works is based on
chiral and SU(4) flavor symmetry and the KSFR relation. The dynamics behind the interaction is assumed to be the
exchange of vector mesons in the Weinberg-Tomozawa term, while the flavor symmetry is broken by using physical
hadron masses.

For our model we follow an approach consistent with heavy quark symmetry which has been used in \cite{juan1}.
Heavy quark symmetry considers the fact that, for infinitely heavy quark masses, the spin interaction of quarks vanishes.
In order to construct the model one mixes SU(4) flavor symmetry with the spin SU(2) symmetry and considers for the
interaction the SU(8) symmetry. Pseudo scalar and vector mesons are treated on equal footing as well as $\oh^+$ and
$\trh^+$ baryons.

The presentation is organized as follows: in the next section we present the mathematical framework on which the SU(8) model
is based and the tools needed for dynamically generating resonances by calculating the scattering T-matrix and identifying
poles on it. Afterwards we present our results and comment on the differences with other works,
identifying a possible candidate for a claimed experimental state. In the last section we make our conclusions.


\section{Framework}

We follow here an $\rm{SU}(8)$ spin flavor scheme. Detailed explanations on the model can be found in reference
\cite{juan1}. Previously used models based on
$\rm{SU}(4)$ symmetry do not
include heavy pseudo-scalar and heavy vector mesons on an equal
footing and this is not justified from the point of view of HQS, which is
the proper spin-flavor symmetry of QCD considering infinitely heavy quark masses.

In $\rm{SU}(8)$, the lowest lying baryons are represented by a
120-plet. The 120-plet breaks into two 20-plets that accommodate the spin $\oh^+$ baryons and 
four 20$^\prime$-plets that are identified with the spin $\trh^+$ baryons. The mesons are represented by a
63-plet plus a singlet. The SU(8) 63-plet contains one SU(4) 15-plet for the pseudo scalar mesons
and three 15-plets plus three singlets that accommodate the vector mesons.

There are four possibilities to construct mesonic and baryonic
hadronic currents that can couple to a singlet in order to construct a
Lagrangian invariant under SU(8) rotations, but only one of these
possibilities reproduces the $\rm{SU}(3)$ Weinberg-Tomozawa Lagrangian
for the light mesons and baryons \cite{juan1}:
\be
\lc_{WT}^{SU(8)}\propto 
((M^\dagger\otimes M)_{63_a}\otimes(B^\dagger \otimes B)_{63})_1 .
\label{eq:wtlag}
\ee
The reduction of this Lagrangian to $\rm{SU}(6)$ reproduces the
Weinberg-Tomozawa Lagrangian used in \cite{juan2}.

In order to break flavor symmetry we use physical masses for the mesons and different meson decay constants:
\be
f_{D_s}=193.7~ {\rm MeV,} \quad f_D=f_{D^*}=f_{D_s^*}=157.4~ {\rm MeV}.
\ee
The $s$-wave tree level amplitudes between two
channels for each $CSIJ$ sector is given by:
\be
V_{ij}^{CSIJ}&=&\xi_{ij}^{CSIJ}\frac{2\sqrt{s}-M_i-M_j}{4f_if_j} 
 \sqrt{\frac{E_i+M_i}{2M_i}}\sqrt{\frac{E_j+M_j}{2M_j}}, \label{eq:pot}
\ee
where $\sqrt{s}$ is the center of mass energy of the system, $M_i$ is
the mass of the baryon in the $i^{th}$-channel, $E_i$ is the energy of
the C.M. baryon in the $i^{th}$-channel, $f_i$ is the decay
constant of the meson in the $i^{th}$-channel and $\xi_{ij}^{CSIJ}$
are coefficients coming from the $\rm{SU}(8)$ group structure of the
couplings. Tables for the $\xi$ coefficients can be found in \cite{meuexo}.

We use this matrix $V$ as kernel to calculate the $T$-matrix:
\be
T^{CSIJ}&=&(1-V^{CSIJ}G^{CSIJ})^{-1}V^{CSIJ},
\ee
where $G^{CSIJ}$ is a diagonal matrix containing the two particle
propagators for each channel. Explicit expressions for the loop
functions are found in \cite{juan3}. With these ingredients one can calculate the T-matrix 
for the scattering of mesons on baryons and identify poles in it, which are interpreted as resonances.

One can also calculate the coupling of the resonance to the different channels by
noting that close to a pole the $T$-matrix can be written as:
\be
T^{CSIJ}_{ij}(z)&=&\frac{g_ig_j}{z-z_{pole}}, 
\ee
where $z_{pole}$ is the pole position in the $\sqrt{s}$ plane and the
$g_k$ is the dimensionless coupling of the resonance to channel $k$. So, by
calculating the residues of the $T$-matrix at some pole, one obtains
the product of the couplings $g_ig_j$.


\section{Results}\label{sec:iii}

Studying the eigen values of the $\xi$ matrices in \Eq{eq:pot} one can know where the interaction is attractive and
therefore how many resonances one might expect. A very rich spectrum is obtained in the $C=-1$ sector. The attractive
SU(3) multiplets in this sector are: for $J=\oh$ two triplets, one $\bar{6}$-plet, two 15-plets and one
$15^\prime$-plet; for $J=\trh$ sector one triplet,
one $\bar{6}$-plet, two 15-plets and one $15^\prime$-plet and for $J=\frac{5}{2}$ only one 15-plet.

Resonances bound by energies of the order of 200-300
MeV in relation to the thresholds of their main channels are expected to be more affected by the approximations made,
since our approach
is based in the Weinberg-Tomozawa term of \Eq{eq:wtlag}. This
Lagrangian is roughly the first order term in a low momentum
expansion, therefore, we expect theoretical uncertainties affecting
our results for such states to be bigger, since higher order
Lagrangians should give sizable corrections. On the other hand some of the states which we obtain are bound by 150
MeV or less \cite{meuexo}. Our results for such states are expected to be more
precise.

Among the resonances we call the
attention to the $S=0$, $I=0$ member of the sextet, generated by the
$N\db$ and $N\dvb$ coupled channel dynamics. This state appears bound by
only 1 MeV, and it is one of our more interesting
predictions. Moreover, it appears as a consequence of treating heavy
pseudo-scalars and heavy vector mesons on an equal footing, as
required by HQS. Indeed, if one looks at the coupled channel matrix,
$\xi_{ij}$ in this sector, one finds the diagonal $N\db
\to N\db$ entry is zero, which means that no resonance in this
sector would be generated if the $N\dvb$ channel is not considered, as
it was the case in Ref. \cite{hofmann}.

One of the poles we obtain for $J=\trh$ has a mass close to 3100 MeV. There is one
experimental claim for an exotic state with $C=-1$ around this mass in \cite{pentaq}.
The state claimed in \cite{pentaq} has been observed in the
decay mode
\be
\Theta_{\bar C}&\rightarrow& N \dvb \rightarrow N \db\pi, \label{eq:dec1}
\ee
Our dynamically generated state has other two
possible decay channels induced by its coupling to channels involving
the $\del (1232)$ resonance, which is not a stable particle. Thus, the
anti-charmed resonance can decay to $\db$ or $\dvb$ plus a virtual
$\del$, which subsequently would decay into a $\pi N $ pair. So, the
dynamically generated state has now another two decay mechanisms apart
from the one in \Eq{eq:dec1}, namely
\be
\Theta_{\bar C}&\rightarrow&\del \db\rightarrow N\pi\db \label{eq:dec2}\\
\Theta_{\bar C}&\rightarrow&\del \dvb\rightarrow N\pi\db\pi.
\ee
The decay in \Eq{eq:dec2} has the same particles in the final state
that in \Eq{eq:dec1}, with the difference that the pion in one case is
coming from the decay of the $D^*$ and therefore has low momentum,
while in the other channel it comes from a $\del$ and may have higher
momentum. The experimental search made in \cite{pentaq} looked only
for pions in order to reconstruct a $D^*$ and may have missed the
other events where the pion comes from a $\del$.


\section{Conclusions}

We have studied the possibility of generating dynamically exotic baryon resonances with $C=-1$ from the meson baryon
interaction. In order to model the meson baryon interaction we have considered an interaction kernel based on the Weinberg-
Tomozawa term with SU(8) spin flavor symmetry which takes into account the heavy quark symmetry of QCD if one considers
infinitely heavy quark masses. From the interaction kernel one calculates the scattering T-matrix and searches this T-matrix
for poles. Poles appearing in the appropriate Riemann sheets are interpreted as resonances.

A very rich spectrum is generated by the model, but the results for many resonances should be considered with caution
since they might appear strongly bound and for such states higher order corrections in the Lagrangian may affect the results.

Some of the states that one obtains from our model cannot be reproduced in models considering only the SU(4) symmetry,
since these models do not treat on equal footing mesons and baryons of different spins.

One of the states obtained in the model could accommodate the experimental claim for an anti-charmed penta quark. We have showed
that this state may couple to more channels than the ones observed experimentally and therefore, further investigation
should help in the identification of this state.


\begin{theacknowledgments}
This work is partly supported by DGI and FEDER funds, under contract
FIS2006-03438, FIS2008-01143/FIS and PIE-CSIC 200850I238 and the Junta
de Andalucia grant no. FQM225. We acknowledge the support of the
European Community-Research Infrastructure Integrating Activity "Study
of Strongly Interacting Matter" (acronym HadronPhysics2, Grant
Agreement n. 227431) under the Seventh Framework Programme of EU.
Work supported in part by DFG (SFB/TR 16, "Subnuclear Structure of
Matter"). L.T. acknowledges support from the RFF program of the University of Groningen.
\end{theacknowledgments}

\bibliographystyle{aipproc}   

\end{document}